\documentclass{MYws-ijmpa2}

\def\bc{\begin{center}}
\def\ec{\end{center}}
\def\beq{\begin{equation}}
\def\eeq{\end{equation}}
\newcommand{\bmath}{\begin{displaymath}}
\newcommand{\emath}{\end{displaymath}}
\newcommand{\beqn}{\begin{eqnarray}}
\newcommand{\eeqn}{\end{eqnarray}}
\newcommand{\beqns}{\begin{eqnarray*}}
\newcommand{\eeqns}{\end{eqnarray*}}
\newcommand{\ba}{\begin{array}{c}}
\newcommand{\bat}{\begin{array}{cc}}
\newcommand{\ea}{\end{array}}

\newcommand{\eqn}[1]{(\ref{#1})}
\newcommand{\be}{\begin{equation}}
\newcommand{\ee}{\end{equation}}
\newcommand{\bel}[1]{\be\label{#1}}

\def\refjl#1#2#3#4#5#6{\bibitem{#1} #2, {\it #3} {\bf #4}, #6 (#5).}
\def\refjla#1#2#3#4#5#6#7{\bibitem{#1} #2, {\it #3} {\bf #4}, #6 (#5); #7.}

\def\etal{{\it et al}.}

\newcommand{\prd}{{\it Phys.~Rev.~D\/ }}
\newcommand{\prc}{{\it Phys.~Rev.~C\/ }}

\newcommand{\npa}{{\it Nucl.~Phys.~A\/ }}
\newcommand{\npb}{{\it Nucl.~Phys.~B\/ }}

\newcommand{\plb}{{\it Phys.~Lett.~B\/ }}
\newcommand{\prl}{{\it Phys.~Rev.~Lett.\/ }}
\newcommand{\epjc}{{\it Eur.~Phys.~J.~C\/ }}
\newcommand{\zpc}{{\it Z. Phys.~C\/ }}
\def\ppnp{{\it Prog. Part. Nucl. Phys.\/ }}
\def\npps{{\it Nucl. Phys.~B (Proc. Suppl.)\/ }}

\def\rpp{{\it Rep. Prog. Phys.\/ }}

\newcommand{\cL}{{\cal L}}

\newcommand{\lsim}{\stackrel{<}{_\sim}}
\newcommand{\gsim}{\stackrel{>}{_\sim}}
\def\eqn#1{(\ref{#1})}

\begin{document}

\markboth{A. Pich} {Tau Physics}

%
\catchline{}{}{}{}{}
%

\title{THEORETICAL OVERVIEW ON TAU PHYSICS\footnote{
Invited talk at the International Workshop on Tau-Charm Physics
(Charm2006), Beijing, China, June 5-7, 2006}}

\author{ANTONIO PICH}

\address{Departament de F\'{\i}sica Te\`orica, IFIC,
Universitat de Val\`encia--CSIC,\\
Edifici d'Instituts d'Investigaci\'o, Apt. Correus 22085,
E-46071 Val\`encia, Spain.\\
Antonio.Pich@ific.uv.es}

\maketitle


\begin{abstract}
Precise measurements of the $\tau$ lepton properties provide
stringent tests of the Standard Model structure and accurate
determinations of its parameters. We overview the present status of
a few selected topics: lepton universality, QCD tests and the
determination of $\alpha_s$, $m_s$ and $|V_{us}|$ from hadronic
$\tau$ decays, and lepton flavor violation phenomena.

\end{abstract}


\section{Lepton Universality}
\label{sec:universality}

In the Standard Model all lepton doublets have identical couplings
to the $W$ boson. Comparing the measured decay widths of leptonic or
semileptonic decays which only differ by the lepton flavor, one can
test experimentally that the $W$ interaction is indeed the same,
i.e. that \ $g_e = g_\mu = g_\tau \equiv g\, $. As shown in
Table~\ref{tab:ccuniv}, the present data verify the universality of
the leptonic charged-current couplings to the 0.2\%
level.\cite{tau02}\cdash\cite{LEPEWWG}    

\begin{table}[ph]
\tbl{Present constraints on $|g_l/g_{l'}|$.}
{\begin{tabular}{@{}ccccc@{}} \toprule
& $B_{\tau\to\mu}/B_{\tau\to e}$  & $B_{W\to\mu}/B_{W\to e}$ & $B_{\pi\to\mu}/B_{\pi\to e}$
\\[3pt]
$|g_\mu/g_e|$ & $1.0000\pm 0.0020$  & $0.997\pm 0.010$ & $1.0017\pm
0.0015$
\\ \colrule
& $B_{\tau\to e}\,\tau_\mu/\tau_\tau$ & $B_{W\to\tau}/B_{W\to\mu}$ &
$\Gamma_{\tau\to\pi}/\Gamma_{\pi\to\mu}$ & $\Gamma_{\tau\to K}/\Gamma_{K\to\mu}$
\\[3pt]
$|g_\tau/g_\mu|$ & $1.0004\pm 0.0022$ & $1.039\pm 0.013$ & $0.996\pm
0.005$ & $0.979\pm 0.017$
\\ \colrule
& $B_{\tau\to\mu}\,\tau_\mu/\tau_\tau$ & $B_{W\to\tau}/B_{W\to e}$
\\[3pt]
$|g_\tau/g_e|$ & $1.0004\pm 0.0023$ & $1.036\pm 0.014$
\\ \botrule
\end{tabular} \label{tab:ccuniv}}
\end{table}

The $\tau$ leptonic branching fractions
and the $\tau$ lifetime  
are already known with a precision of $0.3\%$. It remains to be seen
whether BABAR and BELLE could make further improvements. The $\mu$
lifetime has been measured to a much better precision of $10^{-5}$.
The universality tests require also a good determination of
$m_\tau^5$, which is only known to the $0.08\%$ level. An improved
measurement of the $\tau$ mass could be expected from BES-III,
through a detailed analysis of $\sigma(e^+e^-\to\tau^+\tau^-)$ at
threshold.\cite{Pedro}\cdash\cite{SV:94} 

\section{Hadronic Tau Decays}
\label{sec:hadronic}

The semileptonic decay modes $\tau^-\to\nu_\tau H^-$ probe the
matrix element of the left--handed charged current between the
vacuum and the final hadronic state $H^-$.

For the decay modes with lowest multiplicity,
$\tau^-\to\nu_\tau\pi^-$ and $\tau^-\to\nu_\tau K^-$, the  relevant
matrix  elements  are already  known  from  the  measured  decays
$\pi^-\to\mu^-\bar\nu_\mu$  and  $K^-\to\mu^-\bar\nu_\mu$. The
corresponding $\tau$ decay widths can then be accurately predicted.
As shown in Table~\ref{tab:ccuniv}, the predictions are in good
agreement with the measured values.
%
Assuming universality, these decay modes determine the
ratio\cite{PDG,JOP:06}
\be
\frac{|V_{us}|\, f_K}{|V_{ud}|\, f_\pi} \, = \,\left\{ \bat
0.27618\pm 0.00048 & \qquad
[\,\Gamma(K^-\to\mu^-\bar\nu_\mu)/\Gamma(\pi^-\to\mu^-\bar\nu_\mu)\,
]
\\
0.267\pm 0.005 & \qquad [\, \Gamma(\tau^-\to\nu_\tau
K^-)/\Gamma(\tau^-\to\nu_\tau\pi^-)\, ] \ea\right. .
\ee
The very different accuracy of these two numbers reflects the
present poor precision on $\Gamma(\tau^-\to\nu_\tau K^-)$.

\begin{figure}[pb]\centering
\begin{minipage}{0.4\textwidth}\mbox{}\hskip -1.3cm
\includegraphics[angle=-90,width=6.37cm,clip]{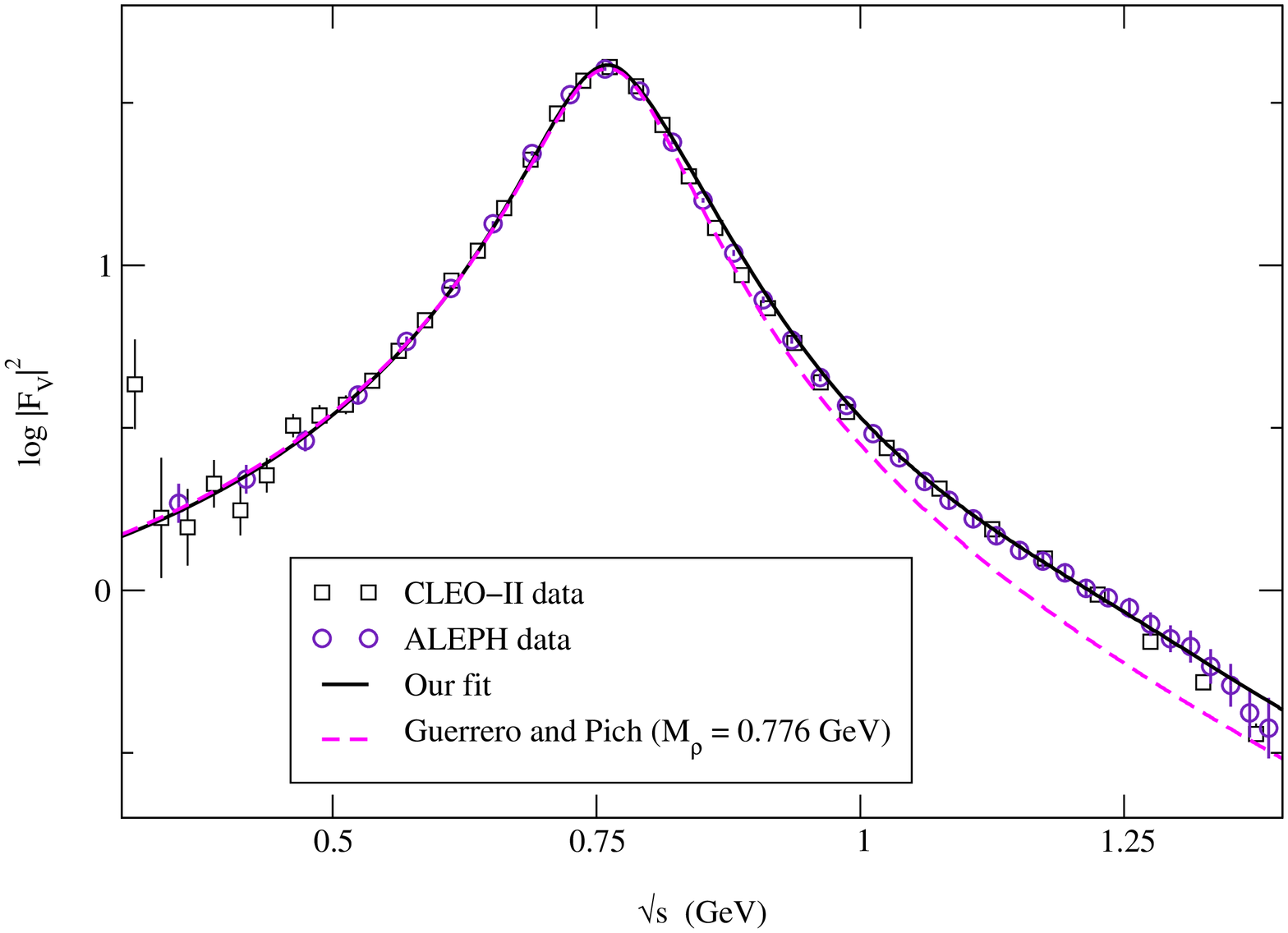}
\end{minipage}
\begin{minipage}{0.4\textwidth}
\includegraphics[angle=-90,width=6.1cm,clip]{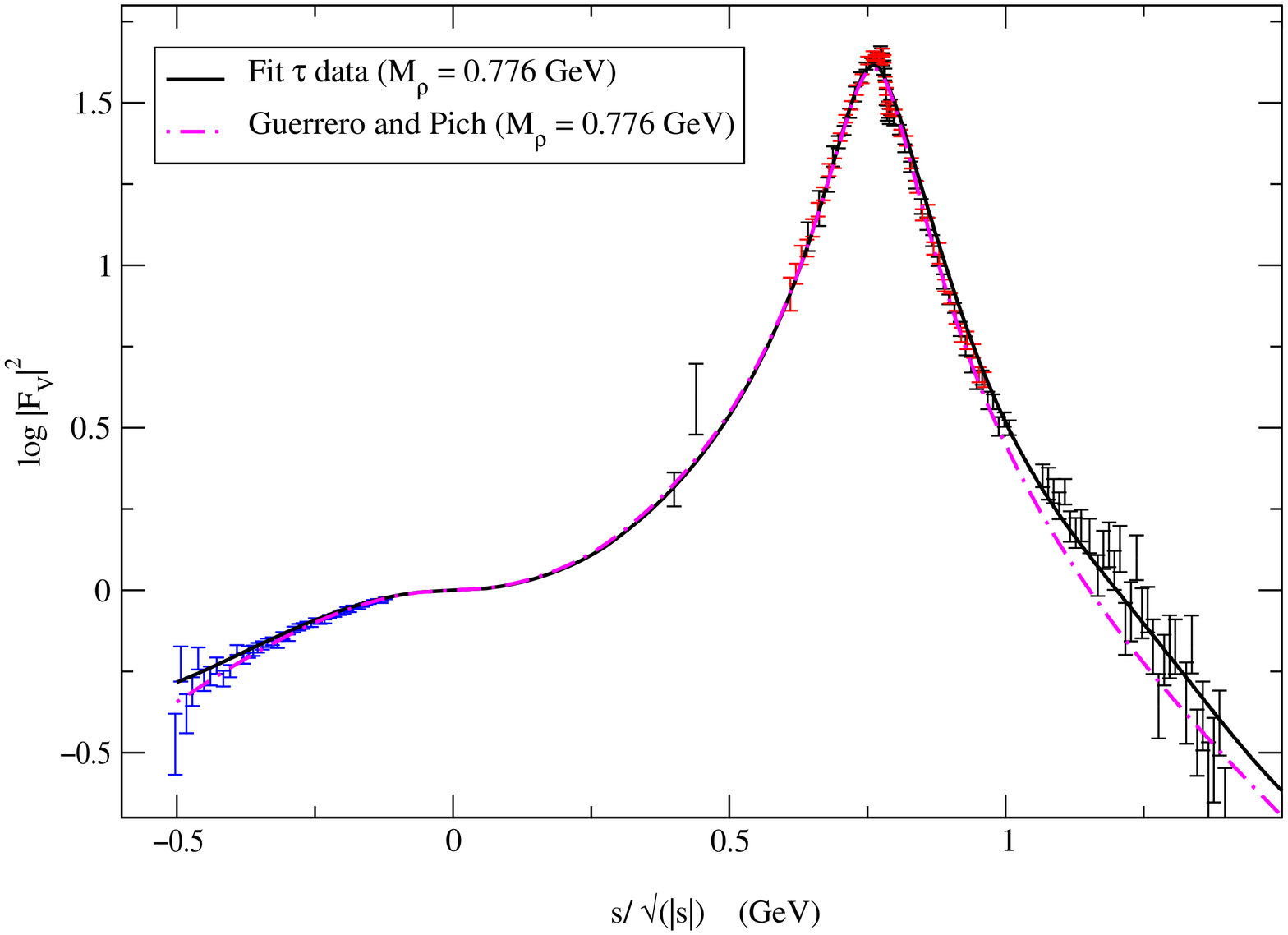}
\end{minipage}
\vspace*{8pt}\caption{Pion form factor from $\tau$
data\protect\cite{ALEPHpiff,CLEOpiff} (left) and $e^+e^-$
data\protect\cite{BA85,AM86} (right), compared with theoretical
predictions.\protect\cite{GP:97,Portoles} The dashed lines
correspond to the result in Eq.~\eqn{eq:PFF_GP}.} \label{fig:pionth}
\end{figure}

For the two--pion final state, the hadronic matrix element is
parameterized in terms of the so-called pion form factor \ [$s\equiv
(p_{\pi^-}\! + p_{\pi^0})^2$]:
\bel{eq:Had_matrix} \langle \pi^-\pi^0| \bar d \gamma^\mu  u | 0
\rangle \equiv \sqrt{2}\, F_\pi(s)\, \left( p_{\pi^-}-
p_{\pi^0}\right)^\mu \, . \ee
A dynamical understanding of the pion form factor can be
achieved,\cite{GP:97}\cdash\cite{RSP:04}
by using analyticity, unitarity and some general properties of QCD,
such as chiral symmetry\cite{GL:85}\cdash\cite{chpt:95}
and the short-distance asymptotic behavior.\cite{EGPR:89,Tempe}
Putting all these fundamental ingredients together, one gets the
result\cite{GP:97}
\bel{eq:PFF_GP} F_\pi(s) = {M_\rho^2\over M_\rho^2 - s - i M_\rho
\Gamma_\rho(s)} \;\exp{\left\{-{s \,\mbox{\rm Re}\left[A(s)\right]
\over 96\pi^2f_\pi^2} \right\}} , \ee
where
\be
A(s) \equiv \log{\left({m_\pi^2\over M_\rho^2}\right)} + 8 {m_\pi^2
\over s} - {5\over 3} + \sigma_\pi^3 \log{\left({\sigma_\pi+1\over
\sigma_\pi-1}\right)}\, ,\qquad
\sigma_\pi\equiv\sqrt{1-4m_\pi^2/s}
\ee
contains the one-loop chiral logarithms
and the off-shell $\rho$ width
is given by\cite{GP:97,DPP:00}
\be \Gamma_\rho(s)\, =\, \theta(s-4m_\pi^2)\,\sigma_\pi^3\, M_\rho\,
s/(96\pi f_\pi^2)\, . \ee
This prediction, which only depends on $M_\rho$, $m_\pi$ and the
pion decay constant $f_\pi$, is compared with the data in
Fig.~\ref{fig:pionth}.\cite{Portoles} The agreement is rather impressive
and extends to negative $s$ values, where the $e^-\pi$ elastic data sits.
%
The small effect of heavier $\rho$ resonance contributions and
additional higher-order (in the Chiral Perturbation Theory and
$1/N_C$ expansions\cite{Tempe}) corrections can be easily included,
at the price of having some free parameters which decrease the
predictive power.\cite{Portoles}\cdash\cite{RSP:04} 
This gives a better description of the $\rho'$ shoulder around 1.2
GeV (continuous lines in Fig.~\ref{fig:pionth}).

More recently, the decay $\tau\to\nu_\tau K\pi$ has been studied in
Ref.~\refcite{JPP:06}. The hadronic spectrum is characterized by two
form factors,
\bel{dGtau2kpi} \frac{d\Gamma_{K\pi}}{d\sqrt{s}} =
\frac{G_F^2|V_{us}|^2 m_\tau^3}{32\pi^3s}
\biggl(1-\frac{s}{m_\tau^2}\biggr)^{\! 2}\Biggl[
\biggl(1+2\,\frac{s}{m_\tau^2} \biggr) q_{K\pi}^3 |F_+^{K\pi}(s)|^2
+ \frac{3\Delta_{K\pi}^2}{4s}\,q_{K\pi}^{\phantom{3}}
|F_0^{K\pi}(s)|^2 \Biggr] \ee
where
$q_{K\pi}^{\phantom{3}}=\frac{1}{2\sqrt{s}}\,\lambda^{1/2}(s,m_K^2,m_\pi^2)$
and $\Delta_{K\pi}= m_K^2-m_\pi^2$. The vector form factor
$F_+^{K\pi}(s)$ has been described in an analogous way to
$F_\pi(s)$, while the scalar component $F_0^{K\pi}(s)$ takes also
into account additional information from $K\pi$ scattering data
through dispersion relations.\cite{JOP:06,JOP} The decay width is
dominated by the $K^*(892)$ contribution, with a predicted branching
ratio Br$[\tau\to\nu_\tau K^*] = (1.253 \pm 0.078)\%$, while the
scalar component is found to be Br$[\tau\to\nu_\tau(K\pi)_{\rm
S-wave}]=(3.88\pm 0.19)\cdot 10^{-4}$.

\begin{figure}[pb]\centering
\includegraphics[width=10cm,clip]{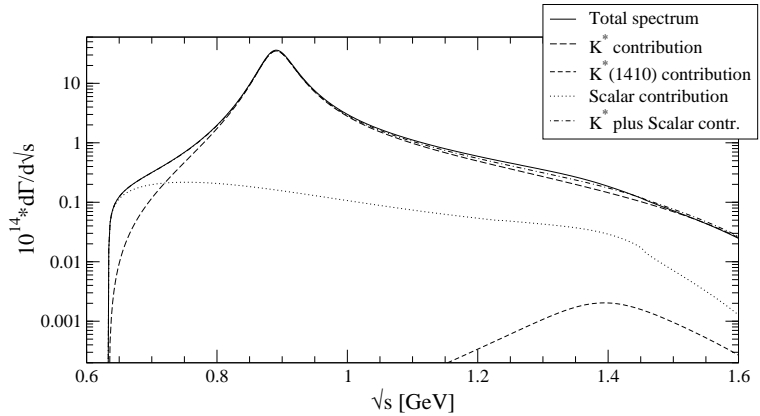}
\vspace*{8pt}\caption{Differential $\tau\to\nu_\tau K\pi$ decay
distribution, together with the individual contributions from the
$K^*(892)$ and $K^*(1410)$ vector mesons as well as the scalar
component residing in 
$F_0^{K\pi}(s)$.\protect\cite{JPP:06}} \label{fig:KpSpectrum}
\end{figure}

The dynamical structure of other hadronic final states can be
investigated in a similar way. The $\tau\to\nu_\tau 3\pi$ decay mode
was studied in Ref.~\refcite{DPP:01}, where a theoretical
description of the measured structure
functions\cite{CLEO3pi}\cdash\cite{ALEPH:05} 
was provided. A detailed analysis of other $\tau$ decay modes into
three final pseudoscalar mesons is in progress.\cite{RPP:06} The
more involved $\tau\to\nu_\tau 4\pi$ and $e^+e^-\to 4\pi$
transitions have been also studied.\cite{EU:02}

\section{The Hadronic Tau Decay Width: $\mathrm{\alpha_s}$}
\label{sec:hadronic_width}

The inclusive character of the total $\tau$ hadronic width renders
possible an accurate calculation of the
ratio\cite{BR:88}\cdash\cite{QCD:94}
%
\bel{eq:r_tau_def}
R_\tau \equiv { \Gamma [\tau^- \rightarrow\nu_\tau
\,\mbox{\rm hadrons}\, (\gamma)] \over\Gamma [\tau^- \rightarrow
\nu_\tau e^- {\bar \nu}_e (\gamma)] }\, = \,
R_{\tau,V} + R_{\tau,A} + R_{\tau,S}\, ,
\ee
using analyticity constraints and the Operator Product Expansion.
One can separately compute the contributions associated with
specific quark currents.
%
%
$R_{\tau,V}$ and $R_{\tau,A}$ correspond to the Cabibbo--allowed
decays through the vector and axial-vector currents, while
$R_{\tau,S}$ contains the remaining Cabibbo--suppressed
contributions.

The theoretical prediction for $R_{\tau,V+A}$ can be expressed
as\cite{BNP:92}
\be R_{\tau,V+A} = 3\, |V_{ud}|^2\, S_{\mathrm{EW}} \left\{ 1 +
\delta_{\mathrm{EW}}' +\delta_{\mathrm{P}} + \delta_{\mathrm{NP}}
\right\} . \ee
%
The factors
$S_{\mathrm{EW}}=1.0194$ and $\delta_{\mathrm{EW}}'=0.0010$ contain
the electroweak corrections at the leading\cite{MS:88} and
next-to-leading\cite{BL:90} logarithm approximation.
The dominant correction ($\sim 20\%$) is the purely perturbative
contribution $\delta_{\mathrm{P}}$, which is fully
known\cite{BNP:92} to $O(\alpha_s^3)$ and includes a resummation of
the most important higher-order corrections.\cite{LDP:92a}

Non-perturbative contributions are suppressed by six powers of the
$\tau$ mass and, therefore, are very small.\cite{BNP:92} Their
numerical size has been determined from the invariant--mass
distribution of the final hadrons in $\tau$ decay, through the study
of weighted integrals,\cite{LDP:92b}
\be R_{\tau}^{kl} \equiv \int_0^{m_\tau^2} ds\, \left(1 - {s\over
m_\tau^2}\right)^k\, \left({s\over m_\tau^2}\right)^l\, {d
R_{\tau}\over ds} \, , \ee
which can be calculated theoretically in the same way as $R_{\tau}$.
The predicted suppression\cite{BNP:92} of the non-perturbative
corrections has been confirmed by ALEPH,\cite{ALEPH:05}
CLEO\cite{CLEO:95} and OPAL.\cite{OPAL:98} The most recent analysis
gives\cite{ALEPH:05}
\bel{eq:del_np} \delta_{\mathrm{NP}} \, =\, -0.004\pm 0.002 \, . \ee

The QCD prediction for $R_{\tau,V+A}$ is then completely dominated
by the perturbative contribution; non-perturbative effects being
smaller than the perturbative uncertainties from uncalculated
higher-order corrections. The result turns out to be very sensitive
to the value of $\alpha_s(m_\tau)$, allowing for an accurate
determination of the fundamental QCD coupling.\cite{NP:88,BNP:92}
The experimental measurement $R_{\tau,V+A}= 3.471\pm0.011$
implies\cite{DHZ:05}
%
\be\label{eq:alpha} \alpha_s(m_\tau)  =  0.345\pm
0.004_{\mathrm{exp}}\pm 0.009_{\mathrm{th}}\, . \ee

The strong coupling measured at the $\tau$ mass scale is
significantly larger than the values obtained at higher energies.
From the hadronic decays of the $Z$, one gets $\alpha_s(M_Z) =
0.1186\pm 0.0027$,\cite{LEPEWWG} which differs from the $\tau$ decay
measurement by more than twenty standard deviations. After evolution
up to the scale $M_Z$,\cite{Rodrigo:1998zd} the strong coupling
constant in \eqn{eq:alpha} decreases to\cite{DHZ:05}
\be\label{eq:alpha_z} \alpha_s(M_Z)  =  0.1215\pm 0.0012 \, , \ee
in agreement with the direct measurements at the $Z$ peak and with a
similar accuracy. The comparison of these two determinations of
$\alpha_s$ in two extreme energy regimes, $m_\tau$ and $M_Z$,
provides a beautiful test of the predicted running of the QCD
coupling; i.e. a very significant experimental verification of {\it
asymptotic freedom}.

\section{Cabibbo--Suppressed Tau Decays: $\mathbf{V_{\! us}}$ and $\mathbf{m_s}$}
\label{sec:ms}

The separate measurement of the $|\Delta S|=0$ and $|\Delta S|=1$ \
$\tau$ decay widths allows us to pin down the SU(3) breaking effect
induced by the strange quark mass,\cite{Davier}\cdash\cite{BChK:05}
through the differences:\cite{PP:99}
\be \delta R_\tau^{kl}  \equiv {R_{\tau,V+A}^{kl}\over |V_{ud}|^2} -
{R_{\tau,S}^{kl}\over |V_{us}|^2} \,\approx\,  24\,
{m_s^2(m_\tau)\over m_\tau^2} \, \Delta_{kl}(\alpha_s) - 48\pi^2\,
{\delta O_4\over m_\tau^4} \, Q_{kl}(\alpha_s)\, . \ee
The perturbative QCD corrections $\Delta_{kl}(\alpha_s)$ and
$Q_{kl}(\alpha_s)$ are known to $O(\alpha_s^3)$ and $O(\alpha_s^2)$,
respectively.\cite{PP:99,BChK:05}
Since the longitudinal contribution to $\Delta_{kl}(\alpha_s)$ does
not converge well, the $J=0$ QCD expression is replaced by its
corresponding phenomenological hadronic
parametrization,\cite{GJPPS:05} which is much more precise because
it is dominated by far by the well-known kaon pole. The small
non-perturbative contribution, $\delta O_4 \equiv\langle 0| m_s \bar
s s - m_d \bar d d |0\rangle
 = -(1.5\pm 0.4)\times 10^{-3}\;\mbox{\rm GeV}^4$,
has been estimated with Chiral Perturbation Theory
techniques.\cite{PP:99}

From the measured moments $\delta R_\tau^{k0}$
($k=0,1,2,3,4$),\cite{ALEPHms,OPALms} it is possible to determine
the strange quark mass; however, the extracted value depends
sensitively on the modulus of the Cabibbo--Kobayashi--Maskawa matrix
element $|V_{us}|$. It appears then natural to turn things around
and, with an input for $m_s$ obtained from other sources, to
actually determine $|V_{us}|$.\cite{GJPPS:05} The most sensitive
moment is $\delta R_\tau^{00}$:
\bel{eq:Vus_formula}
|V_{us}|^2 =
\frac{R^{(0,0)}_{\tau,S}}{\frac{R^{(0,0)}_{\tau,V+A}}{|V_{ud}|^2}-\delta
R^{(0,0)}_{\tau,{\mathrm{th}}}} \, .
\ee
Using $m_s(2~\mathrm{GeV})= (95\pm 20)~\mathrm{MeV}$, which includes
the most recent determinations of $m_s$ from lattice and QCD Sum
Rules,\cite{JOP:06} one obtains $\delta R^{00}_{\tau,{\mathrm{th}}}
= 0.218 \pm 0.026$.\cite{GJPPS:05} This prediction is much smaller
than $R^{(0,0)}_{\tau,V+A}/|V_{ud}|^2$, making the theoretical
uncertainty in \eqn{eq:Vus_formula} negligible in comparison with
the experimental input $R^{(0,0)}_{\tau,V+A}=3.469\pm 0.014$ and
$R^{(0,0)}_{\tau,S}=0.1677\pm 0.0050$.\cite{OPALms} Taking
$|V_{ud}|=0.9738\pm 0.0005$,\cite{PDG} one finally
gets\cite{GJPPS:05}
\bel{eq:Vus_value}
 |V_{us}| =0.2208 \pm 0.0033_{\mathrm{exp}} \pm
0.0009_{\mathrm{th}} = 0.2208 \pm 0.0034\, .
\ee
This result is competitive with the standard determination from
$K_{e3}$ decays, $|V_{us}| =0.2236 \pm 0.0029$.\cite{JOP:06} The
precision is expected to be highly improved in the near future due
to the fact that the error is dominated by the experimental
uncertainty, which can be reduced with the better BABAR and BELLE
data samples. Therefore, the $\tau$ data has the potential to
provide the best determination of $|V_{us}|$.

One can further use the value of $|V_{us}|$ thus obtained in
\eqn{eq:Vus_value} and determine the strange quark mass from higher
$\delta R_\tau^{k0}$ moments with $k\not=0$. One finds in this way
$m_s(m_\tau) = (84 \pm 23)$~MeV, which implies $m_s(2~\mathrm{GeV})
= (81 \pm 22)$~MeV. With future high-precision $\tau$ data, a
simultaneous fit of $m_s$ and $|V_{us}|$ should become possible.

\section{New Physics}

Convincing evidence of neutrino oscillations has been obtained
recently, showing that $\nu_e\to\nu_{\mu,\tau}$ and
$\nu_\mu\to\nu_\tau$ lepton-flavor-violating transitions do
occur.\cite{SNO}\cdash\cite{K2K}  
The non-zero values of neutrino masses constitute a clear indication
of new physics beyond the Standard Model framework.
The simplest possibility would be the existence of right-handed
neutrino components. However, those singlet $\nu_{iR}$ fields would
not have any Standard Model interaction (sterile neutrinos).
Moreover, the Standard Model gauge symmetry
would allow for a right-handed Majorana neutrino mass term of
arbitrary size, not related to the ordinary Higgs mechanism.

In the absence of right-handed neutrino fields, it is still possible
to have non-zero Majorana neutrino masses, generated through the
unique $SU(3)_C\otimes SU(2)_L\otimes U(1)_Y$ invariant operator
with dimension five:\cite{WE:79}
\bel{eq:WE} \Delta \cL\; =\; - {c_{ij}\over\Lambda}\; \bar
L_i\,\tilde\phi\, \tilde\phi^t\, L_j^c \; + \; \mathrm{h.c.}
\stackrel{SSB}{\quad\;\longrightarrow\quad\;} \cL_M = -{1\over 2}\,
\bar\nu_{iL} M_{ij}\, \nu_{jL}^c
 +\mathrm{h.c.} \, ,
\ee
where $\phi$ and $L_i$ are the scalar and $i$-flavored lepton
$SU(2)_L$ doublets, $\tilde\phi \equiv i\,\tau_2\,\phi^*$ and $L_i^c
\equiv \mathcal{C} \bar L_i^t$. After spontaneous symmetry breaking,
$<\phi^{(0)}> = v/\sqrt{2}$, this operator generates a Majorana mass
term for the left-handed neutrinos with $M_{ij} = c_{ij}
v^2/\Lambda$. The Majorana mass matrix mixes neutrinos and
anti-neutrinos, violating lepton number by two units. Clearly, new
physics is called for. Taking $m_\nu\gsim 0.05$~eV, as suggested by
atmospheric neutrino data, one gets $\Lambda/c_{ij}\lsim
10^{15}$~GeV, amazingly close to the expected scale of Gran
Unification.

With non-zero neutrino masses, the leptonic charged current
interactions, involve a flavor mixing matrix $V_L$. Neglecting
possible CP-violating phases, the present data on neutrino
oscillations implies the mixing structure
\be
V_L \; \sim \; \left[  \begin{array}{ccc}
{1\over\sqrt{2}}\, (1+\lambda) & {1\over\sqrt{2}}\, (1-\lambda) & \epsilon \\
-{1\over 2}\, (1-\lambda+\epsilon) & {1\over 2}\,
(1+\lambda-\epsilon) &
{1\over\sqrt{2}} \\
{1\over 2}\, (1-\lambda-\epsilon) & -{1\over 2}\,
(1+\lambda+\epsilon) & {1\over\sqrt{2}} \end{array}\right] ,
\ee
with $\lambda\sim 0.2$ and $\epsilon < 0.2$. Therefore, the mixing
among leptons appears to be very different from the one in the quark
sector.
The number of relevant phases characterizing the matrix $V_L$
depends on the Dirac or Majorana nature of neutrinos. With only
three Majorana (Dirac) neutrinos, the $3\times 3$ matrix $V_L$
involves six (four) independent parameters: three mixing angles and
three (one) phases.

At present, we still ignore whether neutrinos are Dirac or Majorana
fermions. Another important question to be addressed in the future
concerns the possibility of leptonic CP violation and its relevance
for explaining the baryon asymmetry of our universe through a
leptogenesis mechanism.

The existence of lepton flavor violation opens a very interesting
window to improve our understanding of flavor dynamics. The
smallness of the neutrino masses implies a strong suppression of
neutrinoless lepton-flavor-violation processes. However, this
suppression can be avoided in models with other sources of lepton
flavor violation, not related to $m_{\nu_i}$. The present
experimental limits on lepton-flavor-violating $\tau$ decays, at the
$10^{-7}$ level,\cite{LFVbabar,LFVbelle} are already sensitive to
new-physics scales of the order of a few TeV. Further improvements
at future experiments would allow to explore interesting and totally
unknown phenomena.

\section*{Acknowledgments}

I would like to thank the organizers for hosting an enjoyable
conference. This work has been supported in part by the EU
HPRN-CT2002-00311 (EURIDICE), by MEC (Spain, grant
FPA2004-00996) and by GVA 
(Spain, grant ACOMP06/098).



\end{document}